\documentclass[epj]{svjour}

\usepackage{slashed}
\usepackage{graphicx}
\newcommand{\g}{\gamma}
\newcommand{\gmu}{\gamma^\mu}
\newcommand{\gnu}{\gamma^\nu}

\newcommand{\gmunu} {g^{\mu\nu}}
\newcommand{\kmu}{k^\mu}

\newcommand{\kbmu}{\bar{k}^\mu}

\newcommand{\pmu}{p^\mu}

\newcommand{\Nmu}{N^\mu}

\newcommand{\nmu}{n^\mu}

\newcommand{\smu}{s^\mu}
\newcommand{\snu}{s^\nu}
\newcommand{\sbnu}{\bar{s}^\nu}

\newcommand{\be}{\begin{equation}}
\newcommand{\ee}{\end{equation}}
\newcommand{\bea}{\begin{eqnarray}}
\newcommand{\eea}{\end{eqnarray}}
\newcommand{\beal}{\begin{align}}
\newcommand{\eal}{\end{align}}
\newcommand{\bespl}{\begin{split}}
\newcommand{\espl}{\end{split}}

\newcommand{\nsl}{\kern 0.2 em n\kern -0.50em /}
\newcommand{\ksl}{\kern 0.2 em k\kern -0.45em /}
\newcommand{\psl}{\kern 0.2 em p\kern -0.50em /}
\newcommand{\Nsl}{\kern 0.2 em N\kern -0.50em /}
\newcommand{\ssl}{\kern 0.2 em s\kern -0.50em /}
\newcommand{\pbsl}{\kern 0.2 em \bar{p}\kern -0.50em /}
\newcommand{\sbsl}{\kern 0.2 em \bar{s}\kern -0.50em /}
\newcommand{\kbsl}{\kern 0.2 em \bar{k}\kern -0.50em /}
\newcommand{\nbsl}{\kern 0.2 em \bar{n}\kern -0.50em /}
\newcommand{\Nbsl}{\kern 0.2 em \bar{N}\kern -0.50em /}
\newcommand{\Pslash}{\kern 0.2 em P\kern -0.50em /}
\newcommand{\Rslash}{\kern 0.2 em R\kern -0.50em /}

\begin{document}

\title{
Azimuthal asymmetries in exclusive 
four-particle ``Drell-Yan'' 
events.
}

\author{A.~Bianconi}
\institute{Dipartimento di Chimica e Fisica per l'Ingegneria e per i 
Materiali, Universit\`a di Brescia, I-25133 Brescia, Italy, 
\\and 
Istituto Nazionale di Fisica Nucleare, Sezione di Pavia, I-27100 Pavia, 
Italy. \\Email: bianconi@bs.infn.it}

\abstract{
I study hard collisions between unpolarized protons and antiprotons 
where a lepton-antilepton pair is detected in coincidence with a final 
proton-antiproton pair, and no more particles are produced, 
in the regime 10 GeV$^2$ $\ll$ $s$ $\ll$ 1000 GeV$^2$, $M$ 
$>$ 4 GeV, $q_T$ $<$ 3 GeV/c. 
The present work is centered on azimuthal asymmetries. 
Because of momentum conservation, a Boer-Mulders term in the 
momentum distribution of a quark implies a balancing effect in  
the momentum distribution of some spectators. This 
produces azimuthal asymmetries of the final hadrons. 
To analyze this, I have organized a parton-level MonteCarlo 
generator where a standard $\cos(2\phi)$-asymmetry of the dilepton 
distribution is produced, thanks to 
a soft rescattering process between an active quark coming from a 
hadron and a spectator anti-diquark coming from the other hadron. 
This produces $cos(2\phi)$-asymmetries of the final hadron pair. 
Hadron and lepton asymmetries have the same size. 
} 



\maketitle

\section{Introduction}

The rising interest for the so-called 
spin-physics in inclusive hadronic reactions (see 
e.g. \cite{BaroneDragoRatcliffe} or \cite{ferrara08}) 
has among its consequences the 
planning of Drell-Yan experiments\cite{DrellYan,DrellYan2} 
at intermediate energies 
(squared c.m. energy $s$ $\ll$ 1000 GeV$^2$). One is the PANDA 
experiment\cite{panda} where protons and antiprotons collide with variable 
$s$ $\leq$ 30 GeV$^2$ (see 
\cite{PandaPhysicsBook} for an extensive description 
of the experiment goals and methods, and \cite{Bettoni09} for 
a shorter summary).  
This experiment has a broad spectrum of objectives. These include  
unpolarized  Drell-Yan, finalized at a precision 
measurement of the dilepton $cos(2\phi)-$asymmetry  
for $s$ $\approx$ 30 GeV$^2$, and dilepton 
mass $M$ $>$ 2 GeV/c$^2$. 

The $cos(2\phi)-$asymmetry, or $\nu-$asymmetry, was  
first discussed in \cite{LamTung78,LamTung782}. It was found nonzero, with 
size 5-20 \%, in 
$\pi^--$nucleus 
experiments \cite{NA3,NA32,NA10,Guanziroli88,Conway89}, where $s$ 
was between 100 and 600 GeV$^2$ and both $x$ and $\bar{x}$ $\sim$ 
0.1. It was recently found small or zero 
in the $pp$ experiment\cite{E866} at much larger $s$ and 
much smaller average $x$, $\bar{x}$. PANDA will measure it in a 
peculiar large$-x$ regime, with large statistics. 
Both muon pairs 
and electron pairs will be detected. 

Apart for specific kinematics, another peculiarity distinguishes 
PANDA from all 
the previous fixed-target Drell-Yan experiments. Because of 
the multi-purpose nature and design of this apparatus, it 
will detect the class and momentum of 
almost $\underline{all}$ of the reaction products. 

In the past, the most appealing feature of the Drell-Yan measurement 
(in the case of dimuon detection) 
has just been the need not to care other particles apart for the 
$\mu^-\mu^+$ pair. With a thick layer of hadron-absorbing material 
and muon spectrometers downstream, the structure of 
a fixed-target Drell-Yan 
apparatus is relatively simple (see \cite{Matthiae,Kenyon82} 
for reviews). 
For this reason, we have very little information about what is 
produced in Drell-Yan measurements, in coincidence with the 
dilepton pair. The author is  only aware of charge multiplicity 
measurements at ISR\cite{Antreasyan88,Ant2}, 
at the large $\sqrt{s}$ $\sim$ 100 GeV (where a large number 
of particles was produced, of course). 

To analyze what PANDA could find concerning Drell-Yan 
fragments, in a previous work\cite{Nima1} I 
presented a simulation of Drell-Yan events at 
$s$ $=$ 30 GeV$^2$, 
with minimum dilepton mass 2 GeV/c$^2$ and minimum transverse 
momentum 0.8 GeV/c. 
That simulation was performed with 
Pythia-8\cite{Pythia8}. I found some surprising results: 

1) Almost all the events contain one (and only one) 
nucleon-antinucleon pair. 

2) 50 \% of the events $only$ contains a nucleon-antinucleon pair 
(apart for the dilepton, of course). These pairs are equally divided into 
$p\bar{p}$ and $n\bar{n}$ pairs. 

3) In most of the other events the $N\bar{N}$ pair is 
accompanied by not more than two light 
particles (charged pions or photons, possibly from $\pi^0$ decay). 

4) An insertion of the simulated events into the PANDA acceptance 
shows that in the case of ``dilepton $+$ diproton'' events 
with a detected dilepton, 
half of the $p\bar{p}$ pairs will be fully 
detected. This is about 
10 \% of the total rate of detected Drell-Yan events 
in PANDA conditions. 

5) In another 10 \% of detected Drell-Yan events we will only detect 
a single proton (often) or a single antiproton (more rarely), 
but we will be able to identify a ``dilepton $+$ diproton'' 
exclusive event. 

This opens the possibility to analyze 
20 \% of the Drell-Yan events as $exclusive$ events, with some 
interesting associated observables and with the perspective 
of accessible modeling (because of the small number of involved 
particles). 
This work will be centered on an analysis of the 
``dilepton $+$ diproton'' events, including azimuthal asymmetry 
effects. 
Clearly, the standard properties of these events 
may be simulated via Pythia or some other known multi-purpose code 
as I have already done in \cite{Nima1}. 
Here my primary goal is to analyze those unknown properties of the 
proton-antiproton pair, associated with the 
dilepton azimuthal 
$cos(2\phi)-$asymmetry. To this aim, I have organized 
a specific parton-level generator code. 

The extra detected proton-antiproton pair enriches the analysis  with 
two more vectors: $\vec p_P$ and ${\vec p}_{\bar{P}}$. 
These may 
be combined into the two pairs $\vec p_P\pm {\vec p}_{\bar{P}}$. 
In 4-particle events only the relative 
momentum 
$\vec p_P-{\vec p}_{\bar{P}}$ is independent of the dilepton momentum 
$\vec q$ (that is equal to the 
sum of the lepton momenta and to minus the sum of the fragment 
momenta). 
So, here I will focus on the observables related with the 
correlations among the fragment relative momentum, 
the lepton relative momentum and  
$\vec q$.  

Although my main interest is towards PANDA, I will consider a 
hypothetical experiment with $s$ $=$ 100 GeV$^2$ 
instead of 30 GeV$^2$ as in PANDA 
(a possibility in this kinematic range 
could be Drell-Yan at COMPASS\cite{CompassDY}).  
This will allow me to use several simplifying approximations, based 
on the fact that the 
longitudinal momenta of the hadron constituents are for most 
events much larger than the transverse momenta. In the PANDA 
case this is not true, and the analysis would require 
nonstandard devices, and infrared parameters. 

\subsection{Notations}

I will use the letter ``$p$'' to indicate observable momenta 
(leptons, proton, antiproton) and the letter ``$k$'' for the 
momenta of the partons (quark, diquark, antiquark, antidiquark). 
In detail, $p_\mu$ and $\bar{p}_\mu$ indicate lepton momenta,
$k_\mu$ and $\bar{k}_\mu$ (anti)quark momenta, while for the other 
particles a suffix is present. 

The vector $q_\mu$ is the sum of the lepton momenta, and also 
of the quark momenta, and minus the sum of the spectator momenta 
and of the final hadron momenta. I use both $q$ $=$ 
$\sqrt{q_\mu q^\mu}$ and $M$ to  
indicate the same thing, i.e. the invariant mass of the dilepton 
pair. 

The generic letters $\theta$ and $\phi$ indicate all the polar and 
azimuthal angles used in this work. The text and the figure captions 
specify each time which angles I am speaking 
about (suffixes resulted in difficult 
to read figure captions). 

Results will be presented with respect to angles measured in 
two reference frames: 

a) ``hadron collision frame'': the center of mass frame 
of the colliding proton and antiproton, 
with $z$ along the proton direction. The other two axes are 
fixed once and for all, so this frame is the same for 
all events. The fragment momenta are always referred to 
this frame. The lepton momenta are referred to this frame when 
I look for correlations between them and the hadron momenta. 

b) ``co-oriented dilepton c.m. frame'': this is 
an event-dependent frame with all axes parallel to the former one,  
but boosted in such a way to be a center of mass frame for the 
lepton-antilepton pair. This frame is used 
to calculate the azimuthal $cos(2\phi)-$asymmetry of the leptons, 
in a way that is as much similar to the tradition as possible. 
In the present case, the main reason to avoid the 
Collins-Soper\cite{CollinsSoper77} frame and of other similar frames 
with event-dependent orientation of the axes 
is the difficulty in relating lepton and hadron properties. 

Another frame appears in the intermediate stages of the calculation:

c) ``spin frame'': an event-dependent frame,  
whose origin and $z-$axis 
coincide with those of the hadron collision frame,  
and whose $x$ and $y$ axes are random-rotated with flat distribution 
around the $z-$axis. The quark and antiquark spins are projected along 
the $y-$axis of this frame, but no initial or final 
observable quantity refers to it.

\section{General ideas}

I have built a ``Drell-Yan-based'' 
MonteCarlo model: as much as  possible of the knowledge of the 
inclusive Drell-Yan process at partonic level has 
been used here. This includes the leading twist 
factorization properties\cite{CollinsSoperSterman,CSS2,Bodwin,B2}, 
the $x-$distributions and $k_T-$distributions as they have 
been measured in this process and in the required  
regime\cite{Matthiae,Kenyon82}, 
the phenomenology and theory of the inclusive lepton 
$cos(2\phi)$ 
asymmetry\cite{LamTung78,LamTung782,NA3,NA32,NA10,Guanziroli88,Conway89,E866,BoerMulders98,Boer99,BoerBrodskyHwang03}.  
This should not hide the fact that I want to reproduce something 
that is not exactly a Drell-Yan process. In the individual channels 
composing the inclusive Drell-Yan process, higher twist effects may 
be much stronger than in  the overall sum, and 
factorization properties have never been discussed. 

In the hadron-quark-spectator vertex 
I assume a basic diquark form for the spectator, according 
with the widespread quark-diquark model 
used in \cite{JakobMuldersRodriguez97}. 
I will only consider one quark 
flavor (according to \cite{Nima1} 90 \% of the 
proton-antiproton Drell-Yan events derive from  
$u\bar{u}$ annihilations, so this is a good approximation). 

For the hadronization process, I will consider two 
alternative possibilities: 
(i) I will assume that spectator 
momenta may be identified with final hadron momenta, (ii) I 
will assume that the final $q\bar{q}$ pair, needed to allow the 
spectators to form physical hadrons, is created with 
a reasonably distributed relative momentum. 

My main goal is to reproduce 
properties of the final proton-antiproton pair, that have the same 
physical origin of the dilepton azimuthal 
$cos(2\phi)-$asymmetry. 

In the pioneering works by \cite{LamTung78,LamTung782} the 
$cos(2\phi)-$asymmetry is one of the terms appearing in the 
most general form for the dilepton pair 
distribution in unpolarized Drell-Yan.  
Within the scheme given in \cite{BoerMulders98}, this asymmetry 
is rewritten according with the quark-parton model, where it 
is proportional to the product of two terms (one for the quark, one 
for the antiquark) 
of the form $h(x,k_T)\vec s_T \wedge \vec k_T$, where 
$h(x,k_T)$ is the Boer-Mulders function. A commonly 
agreed definition for it may be found in \cite{Trento},  
and some models exist\cite{BoerBrodskyHwang03,GambergGoldstein07,BacchettaContiRadici08,BPM08,Vento09}. 

The $q\bar{q}$ $\rightarrow$ $\gamma^*$ coupling selects 
$q\bar{q}$ pairs with opposite helicities. Combined with the 
Boer-Mulders spin-momentum correlation, this selects $q\bar{q}$ pairs 
with nontrivial space distribution properties. 
The virtual photon 
is able to transfer some of these properties to the dilepton 
relative momentum, resulting in the known $cos(2\phi)-$asymmetry.  

To understand what this could mean for the spectator distribution, 
I trace this path back to its origin: the $cos(2\phi)-$asymmetry 
derives from 
a property of the splitting vertex according to which a quark 
has nonzero $<\vec s_T \wedge \vec k_T>$. 
The spectators must balance this with 
$<\vec s_T \wedge \vec p_{T,spect}>$ $=$ $- <\vec s_T \wedge \vec k_T>$, 
since in the splitting vertex $\vec p_{T,spect} + \vec k_T$ $=$ 
$\vec P_{T,parent}$ $=$ 0. 
This should imply azimuthal asymmetries of the final proton and 
antiproton, for the subset of events where these are observed in 
coincidence with a dilepton pair. 

The spin-momentum correlation picture of \cite{BoerMulders98} 
poses some fundamental problems that have been solved at theoretical 
level\cite{BoerBrodskyHwang03}, but there is some ambiguity about 
how to implement this theory in a parton MonteCarlo model. 
To clarify this, some general premise is useful: 

(i) In a theoretical 
model for a quantum process the expectation values of some variables 
are not obliged to have a physical meaning in the 
intermediate steps. In a MonteCarlo they should better have, 
since all the steps are classical. 
(ii) Loop integrals often present relevant final cancellations. In 
a MonteCarlo a single event is only one of the infinite configurations  
entering a loop. Sometimes cancellations take place anyway over 
a large set of events, sometimes they do not. 
(iii) In a theoretical calculation much is gauge artifact. In a 
MonteCarlo, only physical particles and interactions can appear. 
(iv) Interference processes do not admit a straightforward 
implementation. 
(v) A ``black box'' distribution that turns around problems always 
exists, but one would like to avoid it, as much as possible. 

The product of two 
spin-momentum correlations, each one implying $<s_yk_x>$ $\neq$ 0 
for the quark and the antiquark respectively, does not 
violate general invariance laws. However, each of the two is individually 
unphysical. In \cite{BoerBrodskyHwang03} the ultimate 
reason for the Boer-Mulders spin-momentum correlation 
is searched in initial state interactions 
between active and spectator partons originating from different 
parent hadrons. A nontrivial reduction work shows that it is possible 
to rewrite these effects as a distortion of the 
quark and antiquark initial-state momentum distributions, 
in a factorized format. 

This suggests two possible schemes for a MonteCarlo, that I name 
``strict factorization scheme'' and ``rescattering scheme''. 
In the former one implements the conclusion 
of \cite{BoerBrodskyHwang03}, in the latter the starting point 
of the same work. In other words, in one case one assumes that 
the stationary momentum distribution of 
a quark in a hadron presents unphysical 
spin-momentum correlations,  
in the other case that this is the effect of interactions  
with partons coming from another hadron. 

At leading twist and given some constraints on the rescattering 
interaction, the two schemes lead to 
the same results (for the observable variables discussed in this work), 
for the following two reasons: 

(i) Since the works on leading-twist factorization in 
Drell-Yan\cite{CollinsSoperSterman,CSS2,Bodwin,B2} it is known that gluon 
exchanges in initial state interactions are in large part gauge 
artifacts, or cancel in loops, or may be reabsorbed in 
the distribution functions, or belong to a class of phenomena 
that have no effect in the limit of high collision energy. 
Translated into an implementation of rescattering, 
this means that only exchanges of transverse momentum at a 
fixed scale are admitted (a few GeV/c at most). 
This also implies $k_T$ $\ll$ $k$ for all the 
partons. 

(ii) At leading twist, the $q\bar{q}$ $\rightarrow$ $\gamma^*$ 
$\rightarrow$ $l^+l^-$ cross section 
(with assigned transverse polarizations for $q$ and $\bar{q}$)
is symmetric for quark-antiquark exchange if the quark momenta 
respect $k_T$ $\ll$ $k$ (see eq. \ref{eq:full2W} and the discussion in 
the Appendix). 
As a consequence, an event where an amount 
of transverse momentum is exchanged between a spectator and its  
companion quark, and an event where this momentum is exchanged 
between the same spectator and the 
oppositely coming antiquark, lead to the same observable 
final state variables. 

I will work in the rescattering scheme. It allows for organizing 
a chain of probabilistic steps, none of which is a priori 
unphysical. 

Another problem is posed by the fact that 
rescattering is the effect of an interference process. 
The probability of the considered process 
may be written in the form 
\begin{equation} 
Prob\ \sim\ |(S*P*C\ +\ S*A*C)|^2. 
\end{equation}
Here $S$ is the amplitude 
for the splitting process of the two hadrons and may 
be further factorized as $S(1)S(2)$,  
$P$ describes the 
free propagation of all the partons, 
$A$ is first order rescattering, $C$ is the hard 
$q\bar{q}$ $\rightarrow$ $l^+l^-$ 
process, and ``$*$'' indicates sum/convolution 
with respect to the degrees of freedom of the 
partons in the intermediate stages. 

My assumption is that for any given spin configuration of the 
$q\bar{q}$ pair in the intermediate state 
I may rewrite 
\begin{eqnarray}
|(S*P*C\ +\ S*A*C)|^2\ \rightarrow\ |S|^2*|P+A|^2*|C|^2
\nonumber
\\ = 
|S(1)|^2|S(2)|^2*|P+A|^2*|C|^2
\label{eq:process2}
\end{eqnarray}
and 
substitute $|P+A|^2$ with a physically allowed scattering 
process between physical particles. 

An interaction that is suitable for this aim is the ``scalar $+$ 
spin-orbit'' one, well known in hadron scattering (see 
e.g.\cite{GoldsteinMoravcsik82}). Its effect in the scattering of 
a spin-1/2 projectile by an unpolarized target is to produce a final 
momentum distribution of the form 
$a(k_T) + b(k_T) \vec s_T \wedge \vec k_T$. 
It does not violate any invariance law, 
derives from an interference process between 
scattering and no-scattering amplitudes and modifies the quark 
momentum distribution in the same way as an ``intrinsic'' 
Boer-Mulders function would do. 

What is done here 
is the probabilistic version of the quantum method used 
in \cite{AB_JPG1}, where one rewrites at parton level 
the hadron-hadron spin-orbit interactions. 
The method has some overlap with the 
1-gluon exchange used in perturbative 
models\cite{BoerBrodskyHwang03,GambergGoldstein07,BacchettaContiRadici08,BPM08,Vento09}.

Alternatively, spin-orbit rescattering may be just considered an effective 
way to distort the quark momentum distribution, so to 
introduce a Boer-Mulders term $\propto$ 
$\vec s_T \wedge \vec k_T$ with pre-assigned shape, 
without worrying about 
its physical explanation in more fundamental terms. 

\section{Constraints from the features of inclusive Drell-Yan}

A detailed discussion of the following equations  
may be found in a very extensive form in e.g.\cite{ArnoldMetzSchlegel09}, and 
in more concise and specific 
form in \cite{Conway89,Boer99}, or in many other works. 

Neglecting quark spin effects and assuming one relevant quark flavor, 
the parton model 
$P\bar{P}$ $\rightarrow$ $l^+l^- + X$ cross section 
is 
\begin{equation}
\sigma\ =\ A 
{{u(x,k_T)u(\bar{x},\bar{k_T})} \over{x \bar{x} s}}
{ {H_{\mu\nu}L^{\mu\nu}} \over {q^4} }.
\label{eq:cross_section1}
\end{equation}
Here 
$A$ is a normalization constant, the factor 
$1/x\bar{x}s$ $\approx$ $1/q^2$ 
combines flux and 
phase space factors, $u(x,k_T)$ is the $u-$quark unpolarized 
distribution in an unpolarized proton. 
$L^{\mu\nu}$ is the lepton tensor, $H_{\mu\nu}$ is the $q\bar{q}$ 
tensor, defined so that 
${ {H_{\mu\nu}L^{\mu\nu}} / {q^4} }$ is adimensional. 

The quark and antiquark variables entering 
this equation are 
$x$, $\bar{x}$, $\vec k_T$ and $\vec{ \bar{k_T} }$. From the measurement 
of the lepton momenta $\vec p$ and $\vec{ \bar{p} }$ we may reconstruct 
$x$, $\bar{x}$ 
and $\vec q_T$ $=$ $\vec p_T$ $+$ $\vec{ \bar{p_T} }$ $=$ 
$\vec k_T$ $+$ $\vec{ \bar{k_T} }$. 
The individual quark transverse momenta are not observable, but  
some features of the distribution of 
$\vec k_T$ $-$ $\vec{ \bar{k_T} }$ may be inferred from 
the distribution of $\vec p_T$ $-$ $\vec{ \bar{p_T} }$. 

To introduce quark spin effects in the generator 
I rewrite the above cross section in the 
form 
\begin{equation}
\sigma\ \approx\ \sum_{s_y} \sum_{\bar{s}_y} A 
f(s_y)f(\bar{s}_y)
{{u(x,k_T)u(\bar{x},\bar{k_T})} \over {x \bar{x} s}}
{ {H_{\mu\nu}(s_y,\bar{s}_y)L^{\mu\nu}} \over {q^4}}
\label{eq:cross_section2}
\end{equation}
where 
a random axis $y$ is chosen, and the polarizations $s_y,\bar{s}_y$ are 
random sorted with distribution $f(s_y)$. 

Eq.\ref{eq:cross_section2} 
gives the same results as eq.\ref{eq:cross_section1}, as far as 
the momentum and the spin distributions $u(x,k_T)$ and $f(s_y)$ 
are uncorrelated. The spin-orbit rescattering, or equivalently 
the Boer-Mulders terms, introduces a correlation. In this case 
the phenomenological distribution is 
\begin{eqnarray} 
d\sigma\ \propto { {1+cos^2(\theta)} \over {x\bar{x}s} }
\ \Big(
u(x,k_T)u(\bar{x},\bar{k_T})
\ +\ \nonumber
\\
h(x,k_T)h(\bar{x},\bar{k_T})sin^2(\theta)cos(2\phi)
\Big)\ 
d[cos(\theta)]d\phi.
\label{eq:cross_section3}
\end{eqnarray}
(see sections 4.4 and 5.1 for the definition of $\theta$ and $\phi$). 

The simulation code is parameterized in such a way that the final 
outcome reproduces the features 
of eq.\ref{eq:cross_section3} and of the related 
phenomenology for the dilepton distributions in 
$\pi^-W$ fixed target 
experiments\cite{NA3,NA32,NA10,Guanziroli88,Conway89}.  
We have not data on the  
$cos(2\phi)$ asymmetry in $p\bar{p}$ Drell-Yan, so I 
have to rely on $\pi^-W$ data.  
In both cases we have dominance of valence-valence $u\bar{u}$ 
annihilations. 

\section{MonteCarlo structure} 

The event are sorted at $s$ $=$ 100 GeV$^2$. 
Quarks and leptons are 
considered light-like. With regard to spectators, 
we only need their 3-momentum, so it is not relevant to specify 
their mass. 
All the events are subject to the cutoff $\tau$ $>$ $4^2/100$, 
where $\tau$ $\equiv$ $x\bar{x}$ $\approx$ $M^2/s$.

\subsection{General flowchart}

The chain of steps in the simulation is:

1) The proton with initial momentum $P_0 \hat z$ $\approx$ 
$\sqrt{s}/2 \hat z$ splits 
into a light-like quark and a diquark. The quark has 
3-momentum $[k_x,k_y,xP_0]$ that is sorted randomly: 
$x$ is sorted 
according to a distribution $f(x)/x$; 
$k_x$ and $k_y$ refer to the random-oriented 
spin frame, and are sorted with gaussian 
distribution with center at $k_x$ $=$ $k_y$ $=$ 0. Details 
on the parameters are given later. 
The factor $1/x$ in $f(x)/x$ is needed to take into account 
a factor $1/x\bar{x}$ in the cross section (eq.\ref{eq:cross_section1}).

2) The 3-momentum of the spectator diquark is set to $[-k_x,-k_y,(1-x)P_0]$. 

3) A value $s_y$ $=$ $\pm$ 1 (meaning  
$y-$spin $=$ $\pm \hbar/2$) is sorted for the quark, with probability 
50\% vs 50 \%. At this stage, the sorting of the spin and 
of the other variables are independent. 
The diquark is assumed a scalar one, so it has no spin. As far as 
final state rescattering is neglected, the difference between 
scalar and vector diquarks is not relevant. 

4) The same operations (1, 2, 3) are performed 
for the antiproton, whose initial momentum is $-P_0\hat z$. 

5) Spin-orbit shift of the quark momentum: $k_x$ $\rightarrow$ 
$k_x + s_y \Delta k_x$. The shift $\Delta k_x$ 
is gaussian-distributed, 
with positive average value of $<\Delta k_x>$. 
After this step the quark has $<k_x>$ $=$ $s_y <\Delta k_x>$. 

6) The anti-diquark (i.e. the spectator coming from the 
antiproton) is subject to the opposite 
shift: its initial transverse momentum 
$\bar{k}_{d,x}$ $=$ $-\bar{k}_x$ is shifted to 
$-\bar{k}_x - s_y \Delta k_x$. The result of this and of the 
previous step is that the total momentum of the quark and of the 
anti-diquark is conserved. 

7) Steps (5,6) are symmetrically performed on the $\bar{q}$-diquark 
pair. 

8) The quark and the antiquark variables including spins  
are used to build the hadron tensor. 

9) In the lepton co-oriented c.m. frame (see section 1.1), for $q^2$ 
fixed by the quark and antiquark momenta, the angles specifying the 
direction of the lepton momenta are sorted isotropically. After 
transforming them to the spin frame, the lepton tensor is build. 

10) The lepton and the hadron tensor are contracted, and the  
scalar $W(s_y,\bar{s}_y)$ 
$\equiv$ $H^{\mu\nu}(s_y,\bar{s}_y)L_{\mu\nu}/q^4$ 
is used in an accept/reject 
procedure, where 
the event ``$q\bar{q}$ transformed into 
$l^-l^+$'' is accepted or rejected with probability 
$\propto$  $W(s_y,\bar{s}_y)$  
($W$ does not depend on $q^2$, but only on the relative orientation  
of the momenta). 
If the event is rejected, one restarts from step (1) and all the 
variables have to be sorted again. 

11) In the case of an accepted event, 
the observable momenta (still in the spin frame) 
are transformed to the hadron collision frame 
or to the lepton co-oriented c.m. frame for the data analysis.

\subsection{$\vec k_T$-distributions and $s_xk_y$ distortion}

Here I discuss the sorting of the quark variables. For the antiquark, 
the procedure is the same. 

The initial components of $\vec k_T$ are sorted 
with gaussian distribution,  
with $<k_x>$ $=$ $<k_y>$ $=$ 0, and 
$\sqrt{<k_T^2>}$ $=$ 0.4 GeV/c. 

The shift $\Delta k_x$ is sorted randomly with quasi-gaussian 
distribution with $<\Delta k_x>$ $=$ 0.5 GeV/c and fluctuation 
0.27 GeV/c. ``Quasi-gaussian'' 
means that we have the boundary conditions 
$-1$ GeV/c $<$ $\Delta k_x$ $<$ 2 GeV/c, and that the sorting 
probability tends regularly to zero at these 
limits. 
The value 0.27 GeV/c has the only motivation of reproducing 
the phenomenological data of \cite{NA3,NA32,NA10,Guanziroli88,Conway89}. 

The spin-orbit effect for quarks from a proton 
and antiquarks from an antiproton is here assumed to have the same 
sign of $<(\vec s_T \wedge \vec k_T) \cdot \vec P_{parent}>$.  
This means that  
$<(\vec s_T \wedge \vec k_T) \cdot \hat z>$ is opposite in the two cases, 
since the two parent hadrons are opposite-directed. 

\subsection{$x-$distribution}

The $x-$distribution has the basic form 
$f(x)$ $=$ $x^a(1-x)^b$. Fixed-target Drell-Yan experiments at 
$s$ $\sim$ 100-600 GeV$^2$ used the same parameterization
and gave $a$ $\approx$ 0.5, 
$b$ $\approx$ 2 (see e.g. the appendix of 
\cite{Conway89}). I have adopted these two 
values. 

It must be noted that the lower limit 
$s x\bar{x}$ $>$ (4 GeV)$^2$ weakens the experimental constraints 
on $a$, and decreases the relevance of a precise value for it. 

\subsection{Polarized quark $-$ lepton tensor contraction}

The calculation of this tensor-tensor contraction is reported 
in the Appendix. 
For steps (8,9,10) of the previous flowchart 
I need the probability distribution 

\noindent
\begin{equation}
W(s_y,\bar{s}_y)\ \equiv\ H^{\mu\nu}(s_y,\bar{s}_y)L_{\mu\nu} /q^4
\label{eq:WW} 
\end{equation}
where $L_{\mu\nu}$ is the lepton tensor and $H^{\mu\nu}(s_y,\bar{s}_y)$
is the tensor associated to a quark and an antiquark with assigned 
transverse polarizations. 

Extracting a factor $q$ from each 
vector:
\noindent
\begin{equation}
\kmu\ \equiv\ (q/2) \Nmu, \ \pmu\ \equiv\ (q/2) \nmu, ...
\label{eq:redef1W} 
\end{equation}
I get 
\noindent
\begin{equation}
W_{unpol}\ 
\ =\ 
{1 \over 8}\ [(Nn)(\bar{N}\bar{n})\ +\ (N\bar{n})(\bar{N}n)].
\label{eq:H0bW} 
\end{equation}
and
\noindent
\begin{equation}
W_{s_y\bar{s_y}}\ =\ 
[1 + (s_y\bar{s}_y)]\cdot W_{unpol}\ -\ 
{1 \over 2}s_y\bar{s}_y ( 1 + n_y \bar{n}_y ) 
\label{eq:full2W} 
\end{equation}
with 
$s_y$ $=$ $\pm 1$.
To get to this result, terms with magnitude 
$k_T/q$ have been systematically neglected. So 
eq.\ref{eq:full2W} is a leading twist result. 

The first term in eq.\ref{eq:full2W} is about 
3 times more relevant than the second one in determining the 
overall cross section. 
In the center of mass of the $q\bar{q}$ $\rightarrow$ $l^+l^-$ 
annihilation, where the virtual photon is at rest, 
and the 3-vectors $\vec n$, $\vec N$, etc, are unitary, 
the unpolarized term $W_{unpol}$ 
favors configurations where 
the quark and the lepton directions are aligned, 
according to the known 
``$1 + cos^2(\theta)$'' law. 

In fig.1 I show the distribution $f\big(|cos(\theta)|\big)$ together 
with a fitting curve $\propto$ 
$[1+cos^2(\theta)]$ 
where $\theta$ is the lepton polar angle in the lepton co-oriented 
c.m. frame. In the Collins-Soper frame we should get a precise 
$[1+cos^2(\theta)]-$law, and for large $s$ the $z-$axes of 
two frames coincide.  
Fig.1 shows that we are very close to this condition. 

\begin{figure}[ht]
\centering
\includegraphics[width=9cm]{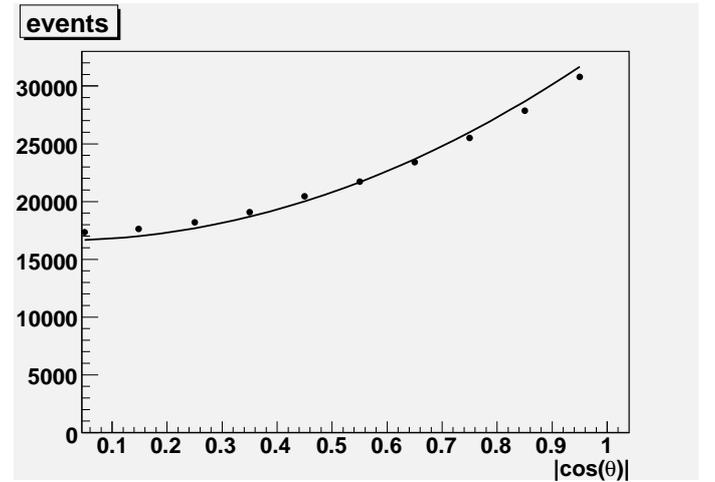}
\caption{Event distribution w.r.t. the polar angle  of the 
positive lepton in the lepton co-oriented c.m. frame. The continuous 
curve is a fit of the form $N$ $=$ $N_0[1+cos(\theta)^2]$. For each 
point, a very small error bar is present although not 
visible, estimated by the standard fluctuation. 
\label{fig1}}
\end{figure}



\section{Results}

I evaluate the measurable asymmetries according 
to the definition 
\begin{equation}
A_\pm\ \equiv\ { {N(+) - N(-)} \over {N(+) + N(-)} }
\label{eq:asymmetry1}
\end{equation}
where 
$N(\pm)$ is the simulated number of events corresponding to 
a positive or negative value of some observable. 
In most of the following 
cases the observable is a factor $cos(2\phi)$, 
where $\phi$ is a lepton or hadron azimuthal angle ranging 
from 0 to $2\pi$. I also present figures with event 
distributions vs $\phi$, and hand-made fits of the 
form 
\begin{equation}
f(N,\beta,\phi)\ \equiv\ N \Big[ 1 + \beta cos(2\phi) \Big]
\label{eq:asymmetry2}
\end{equation}
For a distribution exactly 
coinciding with eq.\ref{eq:asymmetry2},  
the asymmetry calculated according with eq.\ref{eq:asymmetry1} is 
\begin{equation}
A_\pm\ =\ {2 \over \pi} \beta\ \approx\ {2 \over 3} \beta
\label{eq:asymmetry3}
\end{equation}
A typical value that I find is $\beta$ $\approx$ 0.2 corresponding 
to $A_\pm$ 
$\approx$ 0.13. 
 
All the following simulations have been performed once more 
after removing any spin-momentum correlation. 
Because of statistical fluctuation, or of 
uncontrollable numeric systematic effects, these ``vacuum'' 
simulations show asymmetries of 
magnitude 2 \%. So we know that few-percent magnitudes 
cannot be taken seriously given the event numbers 
used in the following. 

\subsection{Behavior of the lepton $cos(2\phi)-$asymmetry}

In this subsection $\phi$ is the angle between the projections 
on the transverse $xy$ plane of the vector $\vec q$ $=$ 
$\vec p+\vec{\bar{p}}$ 
measured in the hadron c.m. frame,  and the vector 
$\vec p-\vec{\bar{p}}$ 
measured in the lepton co-oriented c.m. frame (where it simply 
coincides with $2\vec p$). This is equivalent to the 
ordinary definition of the angle appearing in the 
$cos(2\phi)-$asymmetry, although the traditional procedure 
is to rotate the lepton c.m. frame in such a way that $\phi$ 
is just the angle between the lepton transverse momentum and 
the $x-$axis. Some slightly different choices are normally 
adopted for the  $z-$axis, whose implications have been discussed 
in \cite{BoerVogelsang}. In \cite{Conway89} the data 
have been organized according to different frames, 
and one may appreciate the differences.

In fig.2 I show the $\phi-$distribution for the events with 
$q_T$ between 1.5 GeV/c and 3 GeV/c. 
The fitting curve is 
$1 + 0.23 *cos(2\phi)$ (apart for an  overall normalization). 

\begin{figure}[ht]
\centering
\includegraphics[width=9cm]{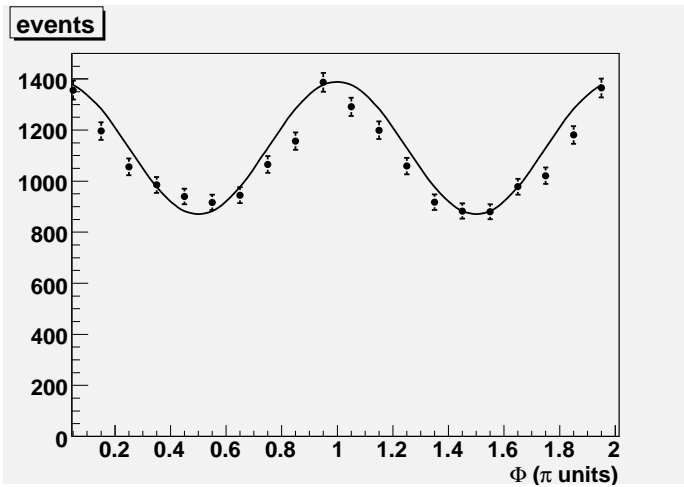}
\caption{$\phi-$distribution of the simulated dilepton events, 
with the transverse momentum of the virtual photon satisfying 
1.5 GeV/c $<$ $q_T$ $<$ 3 GeV/c.  For the definition of $\phi$ 
see the text. The fit is $\propto$ 
$1 + 0.23 *cos(2\phi)$. 
\label{fig2}}
\end{figure}

The applied 
rescattering model does not introduce directly any $x-$dependence 
or $q_T-$dependence: all the quarks are (statistically) 
subject to the same spin-dependent momentum shift. However, 
indirectly a strong $q_T-$dependence and some $x-$dependence 
is produced in the final results. 

In Table 1 I report the event numbers and the asymmetry 
of a set of simulations corresponding to a series of consecutive 
bins in $q_T$. The event numbers in this table 
are proportional to the corresponding cross sections. 
For these event sets a further cutoff has been applied on the 
polar angle: only 
events with $|\theta|$ $<$ 45$^o$ have been counted in this 
statistics. Apart for enhancing the 
$cos(2\phi)-$asymmetry, the second choice is useful to avoid 
the problems described in \cite{AB_torino}, and similar cutoffs 
have been adopted largely in the previously quoted 
experiments because of small-angle and large-angle 
acceptance problems. 

\begin{quote}
\begin{tabular}{|r|r|r|r|}
\hline
Range & Event & Event & Asymmetry \\
(GeV/c) & difference & sum &  \\
\hline
0-0.5 & 163 & 32564 & 0.005  \\
0.5-1 & 1093 & 56314 & 0.02  \\
1-1.5 & 2035 & 33826 & 0.06  \\
1.5-2 & 2061 & 12677 & 0.16  \\
2-2.5 & 872 & 3435 & 0.25  \\
2.5-3 & 208 & 616 & 0.34 \\
\hline
\end{tabular}
\\
\\Table 1: Distribution w.r.t. $q_T$ of dilepton events and of 
their $cos(2\phi)-$asymmetry, with the general 
cutoff $|tan(\theta)| < 1$. For the definition of the $\phi-$angle, 
see the text. 
\end{quote}

\noindent
The chosen parameters of the distorting term produce 
a $cos(2\phi)-$asymmetry with a reasonable magnitude and 
$q_T-$dependence (see the data in \cite{NA3,NA32,NA10,Conway89}, 
and the fit in \cite{Boer99}), at least up to 2.5 GeV/c. 

The experiments \cite{NA3,NA32,NA10,Conway89} 
show a decreasing trend at large $x$, and  
give no information for $x$ below the valence region. 
The proton-proton 
measurement by E866\cite{E866} had larger beam energy and    
smaller average $x$ and $\bar{x}$, and involved one sea active 
quark at least. It 
showed near-zero asymmetries. 

My simulations show a decrease of the asymmetries 
for large and small $x$, $\bar{x}$. It is however difficult to 
disentangle the role of the two.  
To properly analyze the small$-x$ case I should remove the limit $q$ $>$ 
4 GeV/c$^2$ on the dilepton 
mass. This would take me into the region 
$q_T/q$ $\approx$ 1, where I cannot rely on some 
approximations used in building the simulation 
code. 
Also, near the edges of the phase space the physics may be seriously 
affected by constraints that I have not applied here, and by higher twist 
terms. So I would not take 
seriously simulations in the regions where $x$ 
and/or $\bar{x}$ are close to zero or one. In the other regions, the 
asymmetries are roughly $x-$independent. 

\subsection{Proton-antiproton $cos(2\phi)-$asymmetry}

In Table 2, I report final hadron asymmetries and compare them with 
the corresponding lepton asymmetries in two relevant 
kinematic situations. 
For the hadron $cos(2\phi)$ asymmetry, $\phi$ is the angle 
between the difference 
and the sum of the transverse momenta of the proton and of the 
antiproton, 
in the hadron collision c.m. frame (remark: the sum is $-\vec q_T$). 
For the lepton $cos(2\phi)-$asymmetry, the angle is the same 
as in the previous section. Apart for the transverse momentum cuts, 
and for the cutoff 4 GeV/c$^2$ on the dilepton mass, no 
further cuts have been applied to these simulations.  

\begin{quote}
\begin{tabular}{|r|r|r|r|r|}
\hline
Pair & Range & Event & Event & Asymmetry \\
& (GeV/c) & diff. & sum &  \\
\hline
leptons & 1.5 - 3 & 17921 & 133498 & 0.1342 \\
hadrons &         & 17847 & 133498 & 0.1336 \\
\hline
leptons & 0 - 1.5 & 16763 & 974911 & 0.017 \\
hadrons &         & 48550 & 974911 & 0.050 \\
\hline
\end{tabular}
\\
\\Table 2: Compared $cos(2\phi)-$asymmetries for 
leptons and hadrons 
\end{quote}

For the first case of Table 2, the $\phi-$distribution of the 
hadron pairs is reported in fig.3. The fitting curve is 
$1 + 0.22 *cos(2\phi)$ (apart for an  overall normalization). 

\begin{figure}[ht]
\centering
\includegraphics[width=9cm]{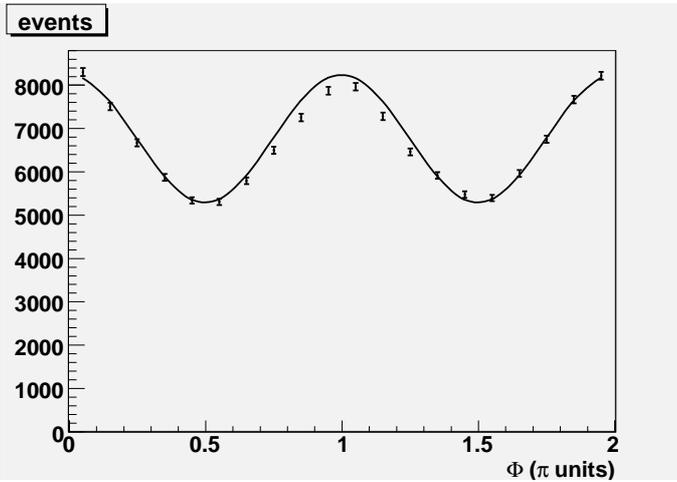}
\caption{Distribution of the proton-antiproton pairs w.r.t. the angle 
$\phi$ between the  
sum and difference of the hadron transverse momenta in the 
hadron collision frame. These events are the same as in Table 2, 
upper part, with cutoff 1.5-3 GeV/c for the total transverse 
momentum and 4 GeV/c$^2$ for the dilepton mass. 
The fit is $\propto$ $1 + 0.22 *cos(2\phi)$. 
\label{fig3}}
\end{figure}

\subsection{hadron-lepton correlations} 

At leading twist the 
hadron and the lepton vectors cannot communicate through 
quantities that are odd in the charge exchange 
on one of the two sides (hadron or lepton). 
I have checked the correlations 
$(p_x\bar{p}_y-p_y\bar{p}_x)_{lept}(p_x\bar{p}_y-p_y\bar{p}_x)_{hadr}$, 
and $(\vec p - \vec{\bar{p}})_{lept}\cdot 
(\vec p - \vec{\bar{p}})_{hadr}$ (transverse components), for which 
I have found nothing meaningful. 

The $\phi-$angle between the relative transverse momentum of the 
leptons and of the hadrons 
could present $\phi-$even systematic behaviors, 
since it 
is the difference between $\phi_{lept}$ and  $\phi_{spect}$ when both 
are measured in the same frame. 
However, the $\phi-$distribution in fig.4 does not show 
reliable systematic deviation from homogeneity. 
I remark that in fig.4 the vertical event scale starts from 
5000. If it started from zero as in the other figures, it would 
appear almost completely flat. The fitting curve is 
$1 + 0.03 *cos(2\phi)$ (apart for an  overall normalization). 
This corresponds to a possible asymmetry about 2 \%. This magnitude 
is within the simulation uncertainties and is not meaningful. 

\begin{figure}[ht]
\centering
\includegraphics[width=9cm]{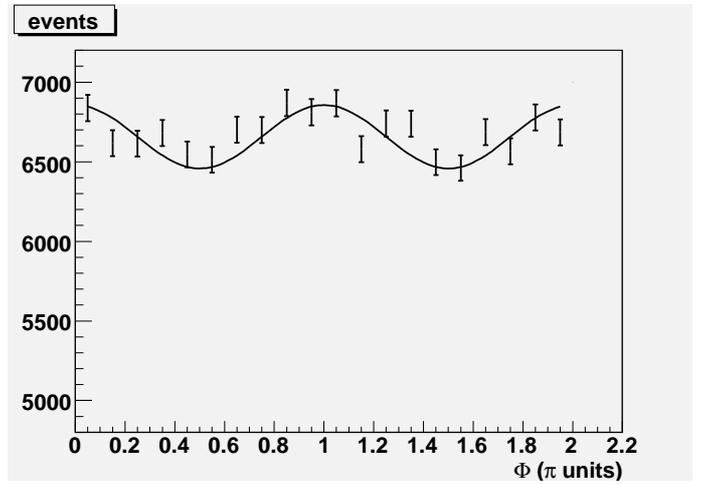}
\caption{Event distribution w.r.t. the angle $\phi$ between the 
relative transverse momentum of the diproton pair and the relative 
transverse momentum of the dilepton pair, both referred to the hadron 
collision frame. Same cuts as in fig.3. The fit is $\propto$ 
$1 + 0.03 *cos(2\phi)$. 
\label{fig4}}
\end{figure}

\subsection{Single hadron - leptons correlations}

A strong correlation is the one between one of the 
hadron fragments (e.g. the proton) and the lepton momentum 
$(\vec p - \vec{\bar{p}})_{lept}$ (transverse components). 

From a theoretical point of view this adds nothing new, 
because the momentum of any of the spectator partons may be written as 
$-\vec q/2$ $\pm$ 
$(\vec p - \vec{\bar{p}})_{spect}/2$, so any observable that is linear 
with respect to the momentum of a spectator is not independent from 
those ones that I have considered previously. 
From the experimenter's point of view, however, to measure a 
proton-dilepton asymmetry in the laboratory frame is easier than 
detecting both a proton and an antiproton (the leptons 
must be detected in any case). According with the analysis 
in \cite{Nima1}, in the specific case of PANDA we have a large 
number of events where it is possible to identify a 
$\mu^+\mu^- p \bar{p}-$event from the detection of 
$\mu^+$, $\mu^-$ and $p$ only.  

In fig.5 I report the distribution of the $\phi-$angle between 
the proton momentum and the difference between the lepton momenta 
(all in the hadronic center of mass frame). The only cutoff is 
$q_T$ $>$ 1.5 GeV/c. 
The corresponding $cos(2\phi)-$asymmetry is 0.14 to be compared 
to the value of the lepton asymmetry that is 0.13. The fit is 
$\propto$ $1+0.22*cos(2\phi)$. 

\begin{figure}[ht]
\centering
\includegraphics[width=9cm]{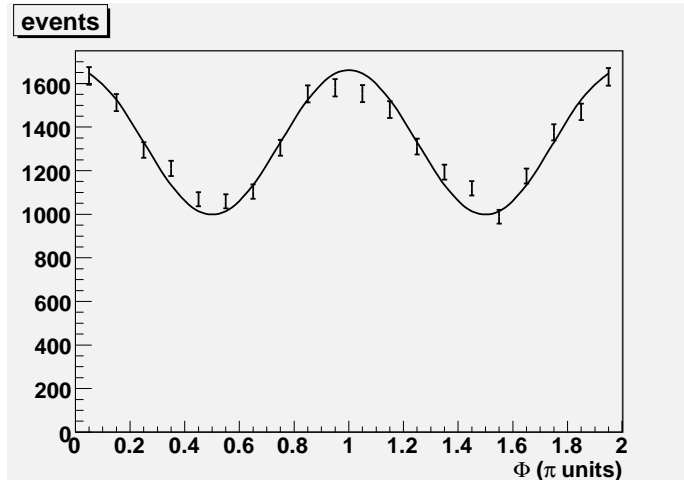}
\caption{Event distribution w.r.t. the angle $\phi$ between the 
proton transverse momentum and the dilepton relative momentum 
in the hadron collision frame. Same cuts as in figs.3 and 4. 
As remarked in the text, the evident asymmetry is just 
another way to measure the 
ordinary dilepton $cos(2\phi)-$asymmetry. The fit is 
$\propto$ $1+0.22*cos(2\phi)$. 
\label{fig5}}
\end{figure}

\section{Higher twist and hadronization effects}

A very simple setup for a realistic hadronization is the one 
used by Pythia in the small phase-space limit of the string 
fragmentation model: the creation of a single 
$q\bar{q}-$pair with a random relative transverse momentum (see 
the detailed description in section 2.4.1 of \cite{Pythia1}, and eq.25 in 
\cite{Pythia2}). I have simulated 
the effect of this on the distribution of the final hadron pair. 

\begin{quote}
\begin{tabular}{|r|r|}
\hline
Gaussian width of  & Asymmetry \\
the relative $P_T$ due &  \\
to hadronization &  \\
\hline
0  & 0.134 \\
\hline
0.35 GeV/c & 0.121 \\
\hline
0.7 GeV/c & 0.091 \\
\hline
\end{tabular}
\\
\\Table 3: Changes of the hadron $cos(2\phi)-$asymmetry associated 
with a relative gaussian-distributed transverse momentum, due to 
the $q\bar{q}$ pair that forms the final hadrons together with 
the diquarks coming from the initial state. 0.134 is the reference 
value from Table 2. 
\end{quote}

\noindent
In other words, my previous simulation relative to table 2 (upper 
line) has been repeated with the 
addition of a gaussian broadening of the relative transverse momentum 
of the final hadron pair. I have tested two cases, where the gaussian 
broadening has width 0.35 and 0.7 GeV/c. The results for the hadron 
$cos(2\phi)$-asymmetry are reported in table 3. 

Since 0.2-0.4 GeV/c is a reasonable range for the width, we see that 
the hadronization process, at least in this simple form, 
does not change the hadron azimuthal asymmetry. 

Some authors (see \cite{Andersson1,Ingelman1,Artru1}) 
have proposed modifications of this basic form of the string model, 
so to include spin-dependent mechanisms in a 2-hadron production stage.  
Probably, this class of processes would more seriously affect the 
hadron asymmetries calculated by me, but they cannot touch the lepton 
asymmetries. 
So, the approximate equality of lepton and hadron 
asymmetries, predicted in this work, is a signature of the absence of 
final state spin-dependent effects. 

The process $p\bar{p}$ $\rightarrow$ 
$\mu^+ \mu^- + p \bar{p}$ 
has chances to be affected by higher-twist effects, since 
it is an exclusive process. 
In a scheme where at least approximate factorization is present 
and higher twist terms are a correction and not the main part of 
the cross section, these effects may be separated into 
initial-state interactions (that link active and spectator partons) 
and  
final-state interactions (not implying direct or indirect 
momentum exchange between spectator and active partons). 

Higher-twist effects 
in the initial state should preserve the approximate equality of 
hadron and lepton asymmetries (although the shape of both could differ from 
what I suggest here). Indeed, the momentum-conservation mechanism 
that is behind the dilepton-dihadron correlation 
is independent of the details of the initial-state processes. 

Summarizing: lepton $cos(2\phi)$-asymmetries that are in disagreement 
with my predictions (fig.2, table 1) are a signature of higher twist 
effects in general, while the inequality of hadron and lepton 
asymmetries is  a signature of specific spin-dependent final state 
effects. 

\section{Conclusions}

This work has considered exclusive 
$p\bar{p}$ $\rightarrow$ 
$\mu^+ \mu^- + p \bar{p}$ 
events at $s$ $=$ 100 GeV$^2$. The main focus 
is (i) the possibility of a lepton $cos(2\phi)-$asymmetry, of the 
same nature of the one observed in inclusive  
$\mu^+ \mu^- + X$ events at the same energy, (ii) the possibility 
of associated asymmetries of the recoiling hadrons. 

The model simulations presented here show that 
$cos(2\phi)$ asymmetries of the leptons in their c.m. frame 
should be accompanied by  
$cos(2\phi)$ asymmetries of the hadrons in the laboratory 
frame. In the latter case, $\phi$ is the angle between the difference 
and the sum of the hadron transverse momenta. 
The size of these asymmetries is the same. 

If the measured asymmetries differ from what I have suggested here, 
this is a signature of higher twist effects (since my predictions 
for the lepton distributions may be read as a re-parameterization 
of the known distributions measured in inclusive Drell-Yan). But, 
if the lepton and hadron asymmetries differ strongly between them, 
this is a more specific signature of final state 
spin-dependent effects.

\appendix
\section{Appendix: Polarized quark$-$lepton contraction 
$H^{\mu\nu}(s_y,\bar{s_y})L_{\mu\nu}$.}

We need the contraction 

\begin{equation}
q^4 W\ \equiv\ H^{\mu\nu}L_{\mu\nu}
\label{eq:W} 
\end{equation}
of the quark-level hadronic and lepton tensors 
in the process 
$q\bar{q}$ $\rightarrow$ $l^+l^-$, where all four particles 
are ultrarelativistic. 
I 
use the shortened  notation for the 
traces 
\begin{equation}
T[..]\ \equiv\ {1 \over 4} Tr[..].  
\end{equation}
I use the definitions of the Berestevskij-Lifsits-Pitaevskij  
book\cite{LL4} (better known as the 4th book of the Landau-Lifsitz 
Course in Theoretical Physics). As the only exception to this, I 
indicate $k_\mu\gamma^\mu$ with the widespread notation $\ksl$ 
instead of using $\hat k$ as was done in that book. 

\subsection{The case of unpolarized quarks}

If 
the quarks are unpolarized, we simply get
\noindent
\begin{eqnarray}
q^4 W_{unpol}\ \equiv\ T[\ksl\gmu\kbsl\gnu]\ T[\psl\gamma_\mu\pbsl\gamma_\nu]\ 
\ =\nonumber \\
=\ 2\ [(kp)(\bar{k}\bar{p})\ +\ (k\bar{p})(\bar{k}p)].
\label{eq:H0} 
\end{eqnarray}
I 
extract a factor $q$ from each 
vector:
\begin{equation}
\kmu\ \equiv\ (q/2) \Nmu, \ \pmu\ \equiv\ (q/2) \nmu, ...
\label{eq:redef1} 
\end{equation}
getting
\noindent
\begin{equation}
W_{unpol}\ 
\ =\ 
{1 \over 8}\ [(Nn)(\bar{N}\bar{n})\ +\ (N\bar{n})(\bar{N}n)].
\label{eq:H0b} 
\end{equation}

In a center of mass frame of the partonic process, the 3-vectors 
$\vec N$, $\vec n$ etc are unitary vectors. In the hadron collision 
frame 
this property is not true, but of course the full tensor contraction 
is frame-independent so $W_{unpol}$ is a scale-independent 
object. The same is valid for the leading-twist terms in the 
spin-dependent $W(s_y,\bar{s}_y)$ that will be calculated below. 

\subsection{polarized quarks (leading twist)}

In the case of a polarized (anti)quark 
\noindent
\begin{equation}
\ksl\ \rightarrow\ \ksl(1-\g^5\ssl)
\label{eq:density} 
\end{equation}
where
$s^\mu$ is the polarization 4-vector for the quark, respecting 
the exact 4-dimensional constraint $(sk)$ $=$ 0. 
If the spin is strictly 
transverse  to the particle motion, then also $\vec s \cdot \vec k$ 
$=$ 0.

In the implementation of a multi-particle MonteCarlo, it would not be 
practical to define a spin quantization axis that accompanies each 
individual quark in its movements, so I have assigned a common 
quantization axis (the $y-$axis) for all spins. Since here 
quarks present nonzero $k_T$ components, this means $O(k_T/q)$ 
helicities. 

The relation between $s^\mu$ and the polarization in a 
rest frame 
$\vec \sigma$ is, for 
massive 
particles,  
\noindent
\begin{equation}
s_0\ =\ \sigma_L |\vec k|/m,\ s_L\ =\ \sigma_L E/m,\ 
\vec s_T\ =\ \vec \sigma_T
\label{eq:spin1} 
\end{equation}
and 
evidently it creates problems for U.R. particles, unless the longitudinal 
component is strictly zero. 
However, when the previous expressions are used to write 
the density matrix eq.\ref{eq:density} in terms of the 
rest frame polarizations, $E/m-$terms cancel and we are left free 
from mass 
singularities: 
\noindent
\begin{equation}
\ksl(1-\g^5\ssl)\ =\ 
\ksl [1 - \g^5 (\pm \sigma_L + \vec \sigma_T \cdot \gamma_T)].
\label{eq:spin2b} 
\end{equation} 
($\pm$ differentiates particles and antiparticles).  
This 
says that a way exists to make calculations without 
meeting mass singularities. 

In the following it will be comfortable to use invariant 
trace formalism. So, it is useful to understand which 
apparently large terms may appear and how to treat them. 

In the product of two $(1-\g^5\ssl)$ density matrices with not exactly 
transverse spins, special care is needed with the terms containing the 
product of one or two longitudinal components. The former give zero trace 
for odd parity of the number of gamma matrixes. The latter modify 
$W_{unpol}$ to 
\noindent
\begin{equation}
W_{h,\bar{h}}\ =\ (1-h\bar{h}) W_{unpol}. 
\label{eq:long} 
\end{equation}
where $h$ and $\bar{h}$ are the average helicities due to the 
longitudinal projections of the 
quasi-transverse spin. The magnitude of these components is 
\noindent 
\begin{equation}
h \sim\ k_y/q, 
\ \bar{h} \sim\ \bar{k}_y/q, 
\label{eq:magnitude1} 
\end{equation}
that leads to $W_{h,\bar{h}}$ $=$ 
$W_{unpol}$ $+$ 
higher twist terms, systematically neglected 
here. 

The full trace for polarized quarks is 
\noindent
\begin{equation}
H_{s,\bar{s}}^{\mu\nu}\ \equiv\ 
T[\ksl\gmu(1 - \g^5\ssl)\kbsl\gnu(1 - \g^5\sbsl)]
\label{eq:full1} 
\end{equation}
\begin{equation}
=\ 
T[\ksl\gmu\kbsl\gnu]\ -\ 
T[\ksl\gmu\ssl\kbsl\gnu\sbsl].
\label{eq:full3} 
\end{equation}

This gives a sum of 15 terms. 
Two of them contain the products $(sk) = 0$ and 
$(\bar{s}\bar{k}) = 0$. We are left with 
\noindent
\begin{eqnarray}
H_{s,\bar{s}}^{\mu\nu}\ =\ 
\Bigg(
[1 - (s\bar{s})]\cdot H^{\mu\nu}_{unpol}\ -\ 
{q^2 \over 2} \{\smu,\sbnu\} \Bigg)\ + 
\nonumber
\\
+\ \Bigg(
(k\bar{s}) \{\kmu,\sbnu \}\ +\ 
(\bar{k}s) \{\kbmu,\snu \}\ -\ 
(k\bar{s})(\bar{k}s) \gmunu
\Bigg)
\label{eq:hfull1} 
\end{eqnarray}
where $\{a_\mu,b_\nu\}$ $\equiv$ $a_\mu b_\nu + a_\nu b_\mu$.  

After contracting the hadron tensor with the lepton tensor, 
the final result consists of many terms. I drop the 
$h\bar{h}-$terms on the ground of eq. \ref{eq:magnitude1}. 
The transverse spin dependent terms contain the product 
$q^4 s_y\bar{s}_y$, and have forms like e.g. 
$q^4 s_y\bar{s_y}N_y\bar{n}_y(\bar{N}n)$. 
I neglect systematically all the products that contain one at 
least among $N_x$, $N_y$, $\bar{N}_x$, $\bar{N}_y$. These terms do 
not reach the magnitude $q^4$, since $N_x$ $\sim$ $k_x/q$ etc.  

I retain all the terms containing 
$(ns)(\bar{n}\bar{s})$, 
$(n\bar{s})(\bar{n}s)$, $(s\bar{s})$ 
(transverse components of $n$ 
and $\bar{n}$ with magnitude $\sim$ $1$ are 
statistically frequent). 

The final result is 
\noindent
\begin{equation}
W_{s_y\bar{s_y}}\ =\ 
[1 + (s_y\bar{s}_y)]\cdot W_{unpol}\ -\ 
{1 \over 2}s_y\bar{s}_y ( 1 + n_y \bar{n}_y ).
\label{eq:full2b} 
\end{equation}

$\\$

{\bf Acknowledgments}

$\\$

The author thanks Alessandro Bacchetta, for long and very useful 
discussions and friendly encouragement. 






\end{document}